\documentclass[aps, pra, onecolumn, 12pt, amsmath, amssymb, amsfonts, notitlepage]{revtex4-1}

\usepackage[english]{babel}
\usepackage[utf8]{inputenc}
\usepackage{graphicx}
\usepackage{tabu}
\usepackage[colorinlistoftodos]{todonotes}
\usepackage{wrapfig}

\def\la#1{\label{#1}}

\newcommand{\myfig}[3]{\resizebox{#1}{#2}{\includegraphics*{#3}}}
\newcommand{\mylab}[3]{\raisebox{#2}[0mm][0mm]{%
\makebox[0mm][l]{\hspace*{#1}#3}}}
\def\aaa{{\sl a}}
\def\bbb{{\sl b}}
\def\ccc{{\sl c}}
\def\ddd{{\sl d}}

\newcommand{\pdd}[2]{\frac{\partial #1}{\partial #2}}
\def\beq{\begin{equation}}
\def\eeq{\end{equation}}
\def\dd{\,{\rm d}}
\def\Rey{\mbox{Re}}

\newcount\ndots
\def\drawline#1#2{\raise 2.5pt\vbox{\hrule width #1pt height #2pt}}
\def\spacce#1{\hskip #1pt}
\def\solid{\drawline{20}{.5}\nobreak}
\def\bdash{\hbox{\drawline{4}{.5}\spacce{2}}}
\def\dashed{\bdash\bdash\bdash\bdash\nobreak}

\def\trian{\raise 0.25pt\hbox{\begin{large}$\vartriangle$\end{large}}\nobreak}
\def\dtrian{\raise 0.25pt\hbox{\begin{large}$\triangledown$\end{large}}\nobreak}
\def\rtrian{\raise 0.5pt\hbox{$\vartriangleright$}\nobreak}
\def\ltrian{\raise 0.5pt\hbox{$\vartriangleleft$}\nobreak}
\def\solidtrian{$\blacktriangle$\nobreak}
\def\solidtri{\raise 0.25pt\hbox{\solidtrian}}
\def\soliddtrian{$\blacktriangledown$\nobreak}
\def\solidrtrian{\raise 0.5pt\hbox{$\blacktriangleright$}\nobreak}
\def\solidltrian{\raise 0.4pt\hbox{$\blacktriangleleft$}\nobreak}
\def\squar{\raise -0.1pt\hbox{$\square$}\nobreak}
\def\circle{\raise -0.5pt\hbox{\begin{Large}$\circ$\end{Large}}\nobreak}

\def\solidcircle{{\Large\raise -0.75pt\hbox{$\bullet$}}\nobreak}
\def\soliddiam{$\blacklozenge$\nobreak}
\def\solidsqua{$\blacksquare$\nobreak}
\def\diam{\raise -1pt\hbox{{\Large $\diamond$}}\nobreak}
\def\smalltimes{\raise 0.6pt\hbox{$\times$}\nobreak}

\def\dasheddot{\bdash$\cdot$\bdash$\cdot$\bdash \nobreak}
\def\dashed{\bdash\bdash\bdash\bdash\nobreak}
\def\linetri{\hbox{\bdash\hspace{-1.3mm}\trian\hspace{-0.8mm}\bdash}\nobreak}

\def\lineltri{\hbox{\bdash\hspace{-1.3mm}\ltrian\hspace{-0.4mm}\bdash}\nobreak}

\def\linesolidtri{\hbox{\bdash\hspace{-1.05mm}\solidtri\hspace{-0.5mm}\bdash}\nobreak}
\def\linesoliddtri{\hbox{\bdash\hspace{-1.3mm}\soliddtrian\hspace{-0.7mm}\bdash}\nobreak}
\def\linesolidrtri{\hbox{\bdash\hspace{-0.9mm}\solidrtrian\hspace{-0.6mm}\bdash}\nobreak}

\def\linesquar{\hbox{\bdash\hspace{-0.9mm}\squar\hspace{-0.3mm}\bdash}\nobreak}
\def\linediam{\hbox{\bdash\hspace{-0.9mm}\diam\hspace{-0.3mm}\bdash}\nobreak}
\def\linesoliddiam{\hbox{\bdash\hspace{-0.9mm}\soliddiam\hspace{-0.3mm}\bdash}\nobreak}
\def\linesolidsquar{\hbox{\bdash\hspace{-0.9mm}\solidsqua\hspace{-0.3mm}\bdash}\nobreak}

\def\linecirc{\hbox{\bdash\hspace{-0.95mm}\circle\hspace{-0.3mm}\bdash}\nobreak}
\def\linesolidcirc{\hbox{\bdash\hspace{-0.9mm}\solidcircle\hspace{-0.3mm}\bdash}\nobreak}

\def\linecross{\hbox{\bdash\hspace{-1.2mm}\smalltimes\hspace{-0.6mm}\bdash}\nobreak}
\def\etal{{\it et al.}}

\usepackage{graphicx}
\usepackage{bm}

\begin{document}

\title{Effect of texture randomization on the slip and interfacial robustness in turbulent flows over superhydrophobic surfaces}

\author{Jongmin Seo}

\author{Ali Mani}

\affiliation{Center for Turbulence Research, Stanford University, Stanford, California 94305, USA}

\begin{abstract}
Superhydrophobic surfaces demonstrate promising potential for skin friction reduction in naval and hydrodynamic applications. Recent developments of superhydrophobic surfaces aiming for scalable applications use random distribution of roughness, such as spray coating and etched process. However, most of previous analyses of the interaction between flows and superhydrophobic surfaces studied periodic geometries that are economically feasible only in lab-scale experiments. In order to assess the drag reduction effectiveness as well as interfacial robustness of superhydrophobic surfaces with randomly distributed textures, we conduct direct numerical simulations of turbulent flows over randomly patterned interfaces considering a range of texture widths $w^+\approx 4-26$, and solid fractions $\phi_s=11\%$ to $25\%$. Slip and no-slip boundary conditions are implemented in a pattern, modeling the presence of gas-liquid interfaces and solid elements. Our results indicate that slip of randomly distributed textures under turbulent flows are about $30\%$ less than those of surfaces with aligned features of the same size. In the small texture size limit $w^+\approx 4$, the slip length of the randomly distributed textures in turbulent flows is well described by a previously introduced Stokes flow solution of randomly distributed shear-free holes. By comparing DNS results for patterned slip and no-slip boundary against the corresponding homogenized slip length boundary conditions, we show that turbulent flows over randomly distributed posts can be represented by an isotropic slip length in streamwise and spanwise direction. The average pressure fluctuation on gas pocket is similar to that of the aligned features with the same texture size and gas fraction, but the maximum interface deformation at the leading edge of the roughness element is about twice larger when the textures are randomly distributed. The presented analyses provide insights on implications of texture randomness on drag reduction performance and robustness of superhydrophobic surfaces. \end{abstract}
\maketitle

\section{Introduction}
\la{sec:intro}
Superhydrophobic surfaces(SHSs) are non-wetting surfaces consisting of hydrophobic chemical coating and micro-nano scale structures that lead to extremely high macroscopic contact angle ($\gtrapprox 150 ^{\circ}$) and small contact angle hysteresis \citep{Rothstein2010, Golovin2016}. When submerged in liquid, micro-structures on SHSs can hold gas pockets which replace the contact area of liquid to solid elements with liquid to gas, causing slippage effect. The macroscopic effect of slippage from bubble mattress has been demonstrated significantly reducing skin friction up to about $20\%$ in laminar flows \citep{Ou2004}. In laminar flows, theoretical investigations of the flow and slippage behavior, depending on the geometric parameters, have been established for the regularly arranged SHSs \citep{Lauga2003, Ybert2007, Davis2010}, and validated \citep{LeeKim2008}. A shear-driven drainage mechanism of failure of slippery surfaces has been studied \citep{Wexler2015a, Liu2016} and a remedy for the shear-driven drainage has been proposed \citep{Wexler2015b}. A great deal of research aimed to test the skin friction reduction effect of SHS in regimes relevant to realistic scenarios, where the overlying flow is turbulent \citep{Daniello2009, Park2014, Bidkar2014, Srinivasan2015, Ling2016}. 

The physical understanding as well as predictions of kinematics of turbulent flows over SHSs have been established by direct numerical simulations (DNSs) of the turbulent channel flows. In DNSs, SHSs are modeled as a surface with a single slip length \citep{Min2004, Fukagata2006, Busse2012}, an idealized air layer with finite thickness \citep{Jung2016}, or patterned slip and no-slip boundaries \citep{Martell2009, Park2013, Jelly2014, Turk2014, Rastegari2015, Seo2015, Seo2016}. Effects of slip on turbulent flow have been first investigated by Min and Kim\cite{Min2004}.  By conducting DNS of turbulent flows over prescribed, single slip length,  Min and Kim\cite{Min2004} showed that the streamwise slip contributes to the upward shift of the whole mean velocity profile by slip velocity in wall unit, $b^+$, and showed that any finite spanwise slip, $b^+_z$, causes downward shift of the outer layer, $\Delta U^+$, due to shrinkage of the buffer layer. Fukagata \etal \cite{Fukagata2006} suggested a phenomenological model for estimation of drag reduction combining effects of streamwise and spanwise slip in a modified form of the log-law. Park \etal \cite{Park2013} conducted DNS of turbulent flows over SHS streamwise ridges for a wide range of texture size and Reynolds numbers and showed that the key parameter determining drag reduction (DR) is $b^+$. Seo and Mani\cite{Seo2016} presented a prediction law for slip length outside of Stokes flow regime as a function of design and flow parameters as $b^+\sim {L^+}^{1/3}/\sqrt{\phi_s}$, where $L^+$ is texture periodicity in wall unit and ${\phi_s}=A_\text{solid}/A_\text{total}$ is solid fraction of the textured surface. Alongside kinematics of SHS in turbulent flows, the mechanisms leading to loss of gas pockets, thus to drag increase, have been studied by Seo \etal \cite{Seo2015} when the gas-liquid interface responds to the pressure fluctuations from overlying turbulent flows. They identified a mechanism of bubble depletion that caused by stagnation of slipping flows at the leading edge of solid roughness elements. Seo \etal \cite{Seo2017} presented an additional breakup mechanism of gas pockets by flow induced capillary waves, when the gas-liquid interfaces dynamically interact with a turbulent flow. The deleterious effect of increasing texture size on the drag observed by these DNS studies is qualitatively consistent with theoretical analysis \citep{Piao2015}, and experimental investigations \citep{Aljallis2013, Bidkar2014, Ling2016}. 

Particularly in economically scalable methods for mass production of SHS, randomly distributed roughness structures are the relevant option in practice \citep{Bidkar2014, Srinivasan2015, Ling2016, Haibao2015, Zhang2015, Hokmabad2016}. In a Stokes flow regime, the relation of slip length to texture size and solid fraction for randomly distributed SHS was first established, analytically by Sbrangalia and Prosperetti\cite{Sbragaglia2007a}. They suggested a heuristic slip length, $b$, in terms of a shear-free hole diameter, $a$, and solid area fraction, $\phi_s$, 
\begin{eqnarray}
b =\frac{8}{9\pi} \frac{(1-\phi_s)}{\phi_s} a,\;\; \phi_s\rightarrow 0. 
\la{eq:sbrpros}
\end{eqnarray}
Samaha \etal \cite{Samaha2011} conducted laminar flow simulations over SHS with randomly distributed textures and showed converged results with different replicas of random distribution of textures for fixed texture size and solid fraction. Samaha \etal \cite{Samaha2011} found that Sbrangalia and Prosperetti's solution is valid for a range of solid fraction $0.1 \le \phi_s \le 0.5$. Bidkar \etal \cite{Bidkar2014} and Ling \etal \cite{Ling2016} showed that the non-dimensional surface roughness in wall unit, $k^+$, was needed to be significantly smaller than the viscous sublayer to obtain successful drag reduction. All previous investigations of turbulent flows over SHS are experimental \citep{Bidkar2014, Srinivasan2015, Ling2016}, while analyses from above-mentioned DNSs only considered geometries with regularly aligned textures. Systematic investigations of texture randomization in DNS can augment quantitative understanding of turbulent flows over the randomly distributed texture. 

In this paper, we investigate the effects of texture randomness in SHS on engineering-level performance when overlaid with a turbulent flow. This study aims to establish the quantitative estimates of drag reduction and interfacial robustness in regimes relevant to practical applications. We explore both kinematic properties as well as pressure loads from DNS data, and the results are compared with SHSs that have the same features but organized in a structured fashion. We will discuss design implications of texture randomness by quantifying performance degradation both in terms of drag reduction and robustness.

\section{Modeling and simulation}
\label{sec:modeling}
\subsection{Governing equations and boundary conditions}
\begin{figure}
\centerline{
\myfig{150mm}{!}{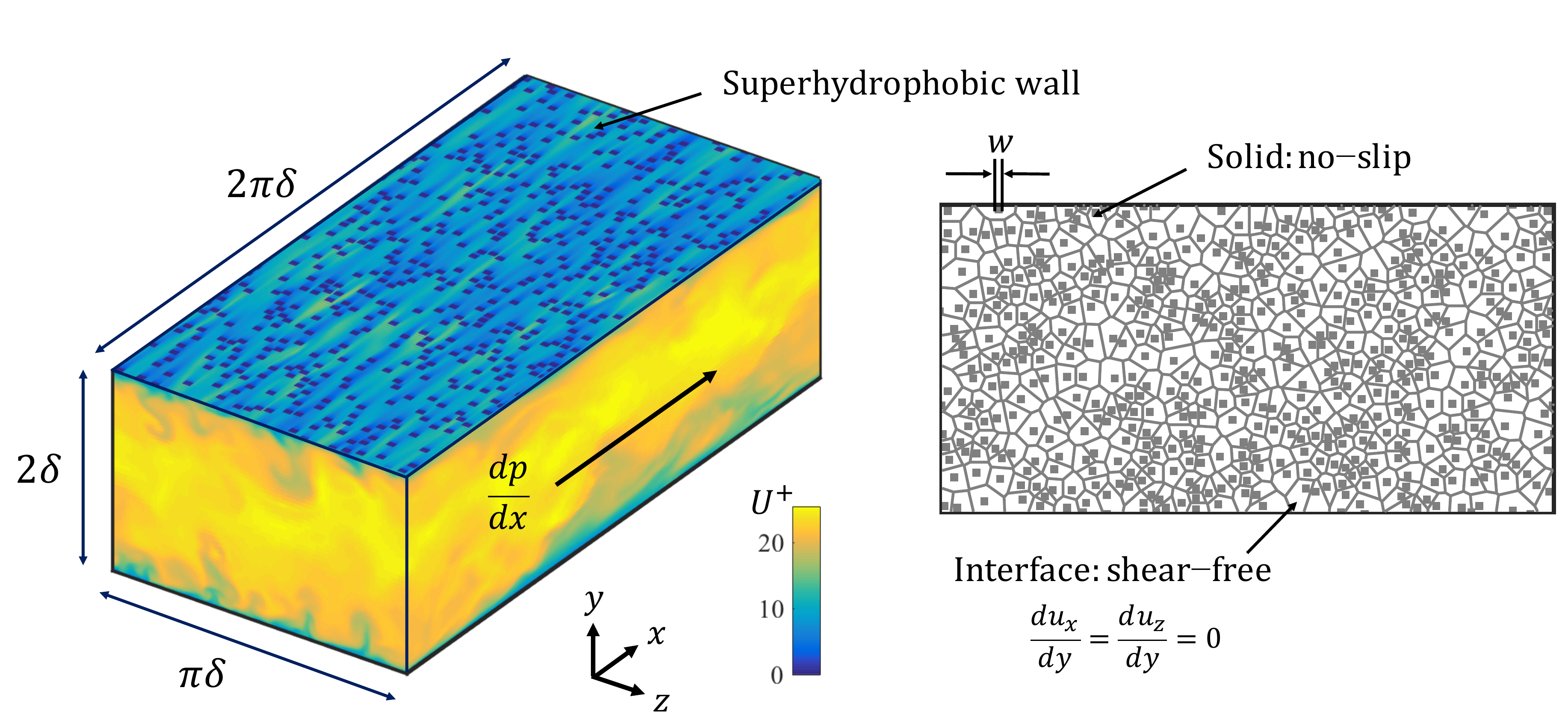}
}
\caption[]{An illustration of periodic turbulent channel with superhydrophobic walls. An instantaneous snapshot of streamwise velocity is plotted. A Voronoi decomposition of surface around textures is shown.}
\la{fig:config}
\end{figure}

We numerically solve the incompressible Navier--Stokes equations for turbulent liquid flows,
\begin{eqnarray}
\nabla \cdot \vec{u} &=& 0,
\label{eq:3Dcont}\\
\pdd{\vec{u}}{t}+\vec{u} \cdot \nabla \vec{u}&=& -\frac{1}{\rho}\nabla p + \nu \nabla^2 \vec{u},
\label{eq:3D_NS}
\end{eqnarray}
in the periodic channel, where $\vec{u}$ is velocity fields, $p$ is liquid pressure, $\nu$ is the kinematic viscosity and $\rho$ is density of liquid.

The channel has top and bottom walls treated with superhydrophobic surfaces consisting of solid roughness elements that entrap gas pockets. The solid-liquid boundary is described by no-slip condition and the gas-liquid interface is treated as shear-free boundary condition as shown in figure \ref{fig:config}. The shear-free boundary condition on the gas-liquid interface is an idealized condition, but is shown to be acceptable for gas-liquid systems involving low viscosity ratios \citep{Schoenecker2014}. The solid roughness elements are randomly distributed over the surface. In current simulations, we do not include the dynamic coupling of the deformation of gas-liquid interface with overlying flows. Instead we consider interface deformation in a post-processing method presented by Seo \etal\cite{Seo2015}, using the wall pressure fields from DNS data. In our analysis, the interface height, $\eta$, measured from the tip of textures $(y=0)$, is quantified solving the linearized Young-Laplace equation,  
\beq
{\sigma}\nabla^2 \eta \approx \Delta P,
\la{eq:Y-L}
\eeq
where $\sigma$ is the surface tension and $\Delta P$ is the pressure difference across the interface, $P_\text{liquid}-P_\text{gas}$. As in Seo \etal\cite{Seo2015}, we assumed the pressure in the gas layer is homogeneous and the mass of the gas is conserved, $\iint{{\eta}\,\dd x \dd z}=0$. 

\subsection{Dimensionless physical parameters}
In our computational modeling, we consider four dimensionless parameters relevant to turbulence, superhydrophobic surfaces and gas-liquid interface. The first parameter is the friction Reynolds-number, $\Rey_\tau=u_\tau \delta /\nu$, which is Reynolds number based on friction velocity $u_\tau=\sqrt{\tau_w/\rho}$, where $\tau_w$ is wall shear, and the boundary layer thickness, $\delta$, equal to channel half height. In Seo \etal\cite{Seo2015}, the effects of SHS on turbulent flows is shown to be insensitive to the Reynolds number, as long as all other parameters are kept constant when normalized by flow inner scale. Guided by these results, in the current study we use $\Rey_\tau=197.5$ in our simulations. 

The second parameter is the width of the texture in viscous unit, $w^+=w\delta/\nu=w/\delta_\nu$. The size of texture characterizes the drag reduction when the solid fraction is fixed for periodically aligned textures \citep{Park2013, Turk2014, Seo2016}. In the present work, we have investigated textures with width $w^+\approx4-26$. We aim to match the smallest texture width $w^+\approx 4$ to that of real superhydrophobic surfaces with texture sizes of $\sim O(10)\mu m$, considering the $\delta_\nu$ in realistic turbulent boundary layer is $\sim O(1)\mu m$. 

Another, texture-related dimensionless parameter is the solid fraction, $\phi_s$. A typical range of $\phi_s$ is $10\%-20\%$ and does not vary by order of magnitude \citep{Bidkar2014, Srinivasan2015, Ling2016}. In the present study we considered $\phi_s=1/9, 1/6$, and $1/4$. Since there is no periodicity in our random texture geometry, we define a nominal texture size, $L^+_n=w^+/\sqrt{\phi_s}$, to compare results with aligned textures. 

The last dimensionless parameter is Weber number $We^+=\rho u_\tau^2 \delta_\nu/ \sigma$. The Weber number can characterize the interfacial robustness in which the momentum from turbulence competes against the surface tension. In our analysis on the interfacial robustness we use $We^+= {10^{-3}-10^{-2}}$, corresponding to a practical application with a typical boundary layer \citep{Seo2015}. 

\subsection{Numerical details}
\begin{table}
\begin{center}
\def~{\hphantom{0}}
\begin{tabular}{ l c  c c c c c c}
\hline
Case   & ${w}^+$& $\phi_s$ &$L^+_n$& $\Rey_\tau$ & $D_x^+$& $D_z^+$ &  $N_x\times N_z \times N_y$\\[3pt]
\hline
R04$\phi_{s25}$      & 4.3 &  1/4 &8.6&    197.5      & 1240.9& 620.5&   1152  $\times$   576  $\times$  128\\
R04$\phi_{s16}$      & 4.3 &  1/6 &10.5&    197.5      & 1240.9& 620.5&   1152  $\times$   576  $\times$  128\\
R04$_{1}$               & 4.3 &  1/9 &12.9&    197.5      & 1240.9& 620.5&   1152  $\times$   576  $\times$  128\\
R04$_{2}$              & 4.3 &  1/9 &12.9&    197.5      & 1240.9& 620.5&   1152  $\times$   576  $\times$  128\\
R08           	      & 8.6  &  1/9&25.8&    197.5    & 1240.9 & 620.5&  576  $\times$   288  $\times$  128  \\
R13           	      & 12.9  &1/9&38.7  &    197.5   & 1240.9 & 620.5 &   384  $\times$   192  $\times$  128 \\
R26$_{1} $             & 25.8   &1/9&77.4 &    197.5     & 1240.9& 620.5&   384  $\times$  192  $\times$  128 \\
R26$_{2}$             & 25.8   &1/9&77.4 &    197.5     & 1240.9& 620.5&   384  $\times$  192  $\times$  128 \\
R26$_{3}$             & 25.8   &1/9 &77.4&    197.5     & 1240.9& 620.5&   384  $\times$  192  $\times$  128 \\
R26$_{4}$             & 25.8   &1/9&77.4&    197.5     & 1240.9& 620.5&   384  $\times$  192  $\times$  128 \\
\end{tabular}
\caption[]{Simulation parameters. $w^+$ is the texture width, $\phi_s$ is solid area fraction, and $\Rey_\tau$ is the friction Reynolds-number. $L^+_n$ is a nominal texture size, $L^+_n=w^+/\sqrt{\phi_s}$. Domain size in viscous unit is $D_x^+$ and $D_z^+$ for streamwise and spanwise directions. The number of grid points are $N_x$, $N_z$, and $N_y$, for streamwise, spanwise, and wall-normal directions respectively. Subscripts represent the replicated cases with different texture arrangements for the same texture parameters. 
}
\la{tab:pressure}
\end{center}
\end{table}
The three-dimensional Navier-Stokes equations are numerically solved with the code of Seo \etal\cite{Seo2015} using a second-order finite-difference method on a staggered mesh. The channel size and number of grid points are described in table \ref{tab:pressure}. Independence of turbulence statistics on size of grid and domain is well verified \citep{Seo2015}. In addition to the patterned slip and no-slip simulations, we conducted DNS over homogenized slip surfaces with a uniform, isotropic slip length, $b^+$ as used in Min and Kim\cite{Min2004}. The DNSs with homogenized slip model use the slip lengths obtained from DNS of flows over surfaces with patterned boundary conditions. In the homogenized simulations, number of grid points is down to 192($x$) $\times$ 192($z$) $\times$ 128($y$). The fractional-time advancement scheme uses the second-order Adams-Bashforth scheme for the non-linear and wall-parallel diffusion terms, and the second-order Crank-Nicholson scheme for the wall-normal diffusion terms \citep{KimMoin1985}. 
The simulations were run for at least 35$\delta/u_\tau$ and only the snapshots after $t=10\delta/u_\tau$ were used for statistical sampling to exclude the initial transient signals. The flow is driven by a time-constant mean pressure gradient to ensure the pre-specified $w^+$. 

\subsection{Texture arrangement and repeatability of statistics}
\begin{figure}
\centerline{
\myfig{155mm}{!}{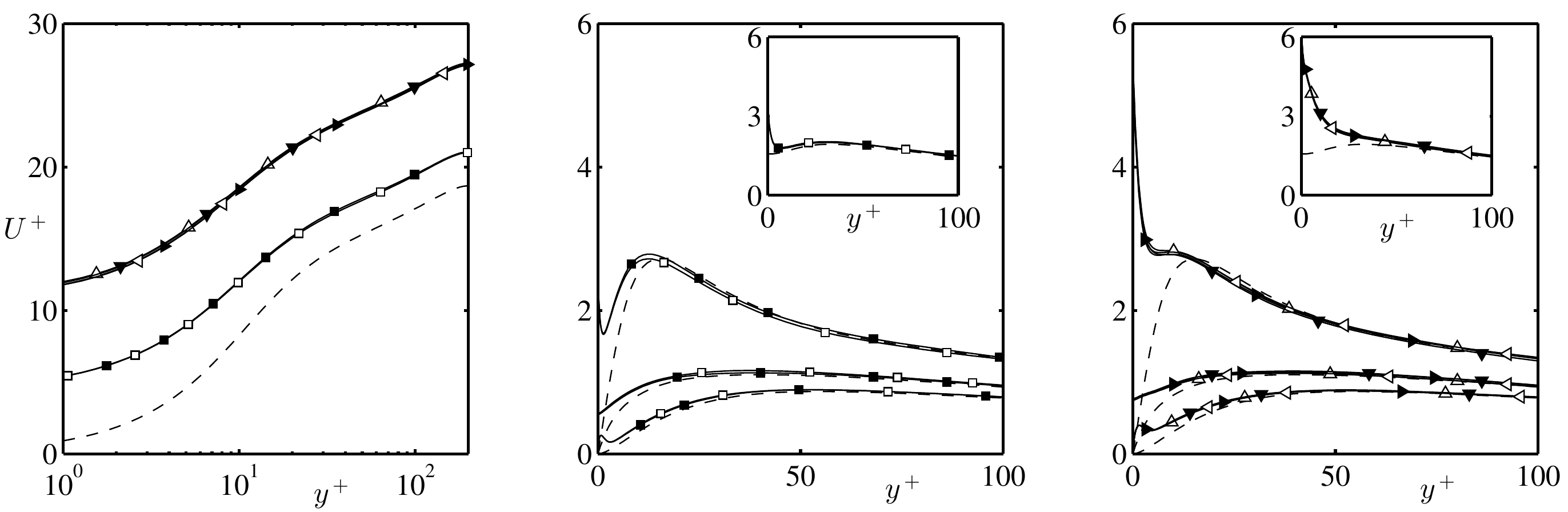}
\mylab{-160mm}{48mm}{(\aaa)}
\mylab{-107mm}{48mm}{(\bbb)}
\mylab{-56mm}{48mm}{(\ccc)}
\mylab{-90mm}{25mm}{\footnotesize{$u_x'^+$}}
\mylab{-88.5mm}{8.mm}{\footnotesize{$u_y'^+$}}
\mylab{-98mm}{15.8mm}{\footnotesize{$u_z'^+$}}
\mylab{-82mm}{39mm}{\footnotesize{$p'^+$}}
\mylab{-45mm}{26mm}{\footnotesize{$u_x'^+$}}
\mylab{-42.5mm}{7.5mm}{\footnotesize{$u_y'^+$}}
\mylab{-48mm}{16.5mm}{\footnotesize{$u_z'^+$}}
\mylab{-31mm}{39mm}{\footnotesize{$p'^+$}}
}
\caption[]{\la{fig:repeatability_small}
Turbulence statistics with different post arrangements with $\phi_s=1/9$ at $Re_\tau\approx200$. (\aaa) Mean streamwise velocity profile; (\bbb) $w^+\approx4$, rms velocity and pressure fluctuations; (\ccc) $w^+\approx26$, rms velocity and pressure fluctuations. \dashed: smooth wall.  \linesolidsquar, R04$_{1}$;  \linesquar, R04$_{2}$;  \linesoliddtri, R26$_{1}$;  \linetri, R26$_{2}$;  \linesolidrtri, R26$_{3}$; \lineltri, R26$_{4}$. 
}
\end{figure}
We randomly distribute posts in the computational domain and avoid overlapping using the algorithm used by Samaha \etal\cite{Samaha2011}. Specifically, the $i$'th post is located randomly in the domain then if any of distance $r_{ij}=\sqrt{r_i-r_j}, j=(1,2,...i-1)$, is less than $2\sqrt{2}w$, the allocation of the post is ignored and a new random location is tested. In this notation, $r_i=(x_i,z_i)$ is the location of $i$'th number of post, and $r_j$ is the positions of other posts assigned before the $i$'th post. This procedure is repeated until the total number of posts matches the targeted solid fraction.

To assess the variability of results due to different random ensembles, we consider multiple realizations of random textures formed by the same texture parameters, $w^+$ and $\phi_s$. Two replicas for $w^+\approx 4$ and $\phi_s=1/9$ are generated and all statistically averaged flow quantities (e.g. mean velocity profile, turbulence intensities) from the two samples collapse within $2\%$ difference as shown in figure \ref{fig:repeatability_small}. Likewise, we additionally examine the repeatability of the statistics for larger textures with $w^+\approx 26$. Four replicas are generated with different texture arrangements. Turbulent statistics show that all different texture arrangements produce results within 3$\%$ difference throughout the channel as shown in figure \ref{fig:repeatability_small}. This implies that our domain had sufficiently large areas for statistically converged mean friction, even when a single realization of these surfaces is used.

\section{Results and discussion}

\la{sec:results}

\subsection{Slip length and drag reduction}

\begin{figure}
\centerline{
\myfig{166mm}{!}{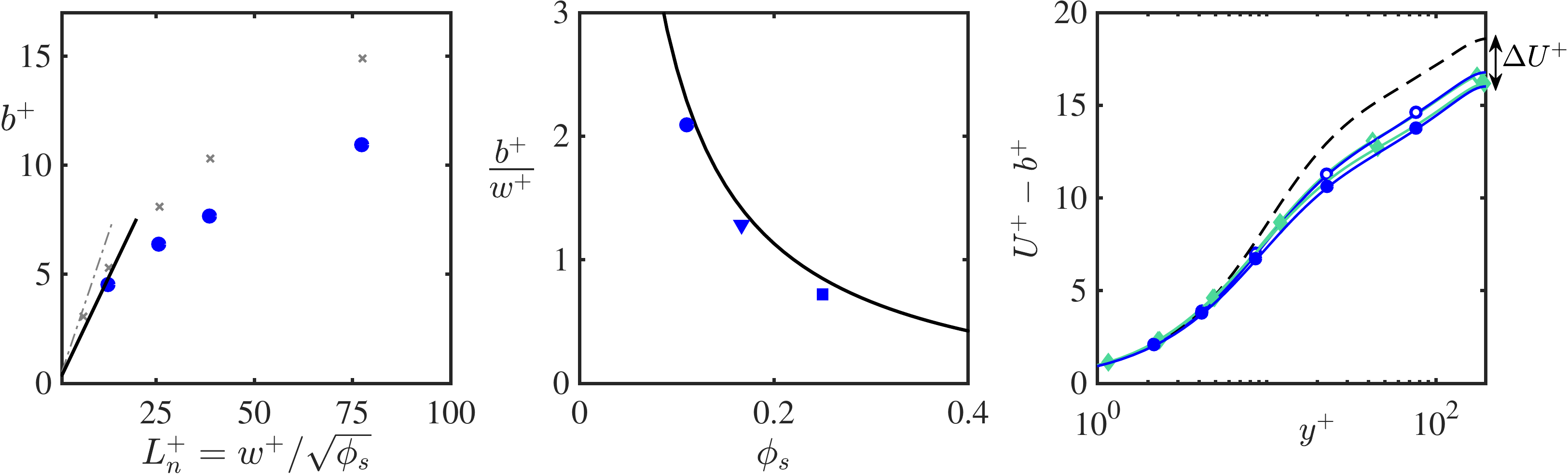}
\mylab{-170mm}{48mm}{(\aaa)}
\mylab{-115mm}{48mm}{(\bbb)}
\mylab{-64mm}{48mm}{(\ccc)}
}
\caption[]{  
(\aaa) Slip length, $b^+$, versus texture size, $L^+_n=w^+/\sqrt{\phi_s}$, with $\phi_s=1/9$. Symbols are DNS solutions; \solidcircle: randomly distributed posts, \smalltimes: aligned posts \citep{Seo2016}.  Lines are Stokes flow solutions; \solid: Equation (\ref{eq:sbrpros}) by \cite{Sbragaglia2007a}; \dasheddot: Ybert \etal \cite{Ybert2007} (\bbb) Slip length, $b^+$, normalized by texture width, $w^+$, versus solid fraction, ${\phi_s}$, at $w^+\approx 4$. \solidcircle: R04$_{1}$, \soliddtrian: R04$\phi_{s16}$, \solidsqua: R04$\phi_{s25}$,  \solid: \cite{Sbragaglia2007a} (\ccc) Mean velocity profiles subtracted by slip velocity for $\phi_s=1/9$,  \linecirc, Patterned slip with randomly distributed posts, R04$_{1}$; \linesolidcirc, Patterned slip, R26$_{1}$; \linediam, Homogenized slip using the slip length from R04$_{1}$, \linesoliddiam, Homogenized slip using the slip length from R26$_{1}$. \dashed: smooth-wall.
}
\la{fig:sliplength}
\end{figure}

We first analyze the impact of geometric randomness on slip length as compared to that of aligned posts. For each geometric configuration, we fix $L^+_n=w^+/\sqrt{\phi_s}$ (as well as $w^+$). The slip lengths of randomly distributed SHS textures with widths $w^+\approx 4-26$, and a fixed solid fraction $\phi_s=1/9$, is plotted in terms of effective texture size in figure \ref{fig:sliplength}(\aaa). In the small texture size limit, the slip length from the DNS data matches with the heuristic model by Sbrangalia and Prosperetti \cite{Sbragaglia2007a}. The validity limit of Stokes flow solution is up to $L^+_n\approx13$ for the randomly distributed post case, while the validity of Stokes flow solution for the aligned post case was limited up to $L^+_n\approx 10$. Figure \ref{fig:sliplength}(\bbb) shows that for $w^+\approx4$ the dependence on $\phi_s$ matches the Stokes flow prediction by Sbrangalia and Prosperetti \cite{Sbragaglia2007a}. In the large effective texture size limit, the slip length increases non-linearly as previously observed for periodic geometries \citep{Seo2016}. In comparison with the DNS data from aligned posts, the randomly distributed posts produce approximately $30\%$ less slip length for the same nominal texture size $L^+_n$. 

We compare the mean velocity profiles obtained from simulations of homogenized isotropic slip length models and that of the corresponding DNS of randomly distributed posts in figure \ref{fig:sliplength}(\ccc). The mean velocity profile from homogenized slip simulation with isotropic slip length matches well with that from patterned slip DNS in the entire channel. 
The modification of mean velocity profiles for SHSs due to shrinkage of the buffer layer is measured by calculating the velocity deficit $\Delta U^+=U_{b,smooth}-(U_{b,SHS}-b^+)$ at the outermost location from the wall, $y=\delta$ as shown in figure \ref{fig:sliplength}(\ccc). The excellent match of the velocity profile of homogenized model with that of patterned slip and no-slip DNS implies that the estimation of $\Delta U^+(b^+)$ from isotropic slip length model, such as $\Delta U^+=4-(1-(4/(4+b^+_z))$ in Busse and Sandham\cite{Busse2012}, can be used to predict drag reduction for SHS with randomly distributed texture. The validity of homogenized isotropic slip length model covers a wide range of texture size up to $L^+_n\approx77$. In contrast, for surfaces with aligned textures a previous analysis by Seo and Mani\cite{Seo2016} showed that the the validity is limited to texture size of $L^+_n\lesssim10$.

\subsection{Pressure fluctuation and interfacial robustness}

\label{sec:interface_random}
 \begin{figure}
\centerline{
\myfig{175mm}{!}{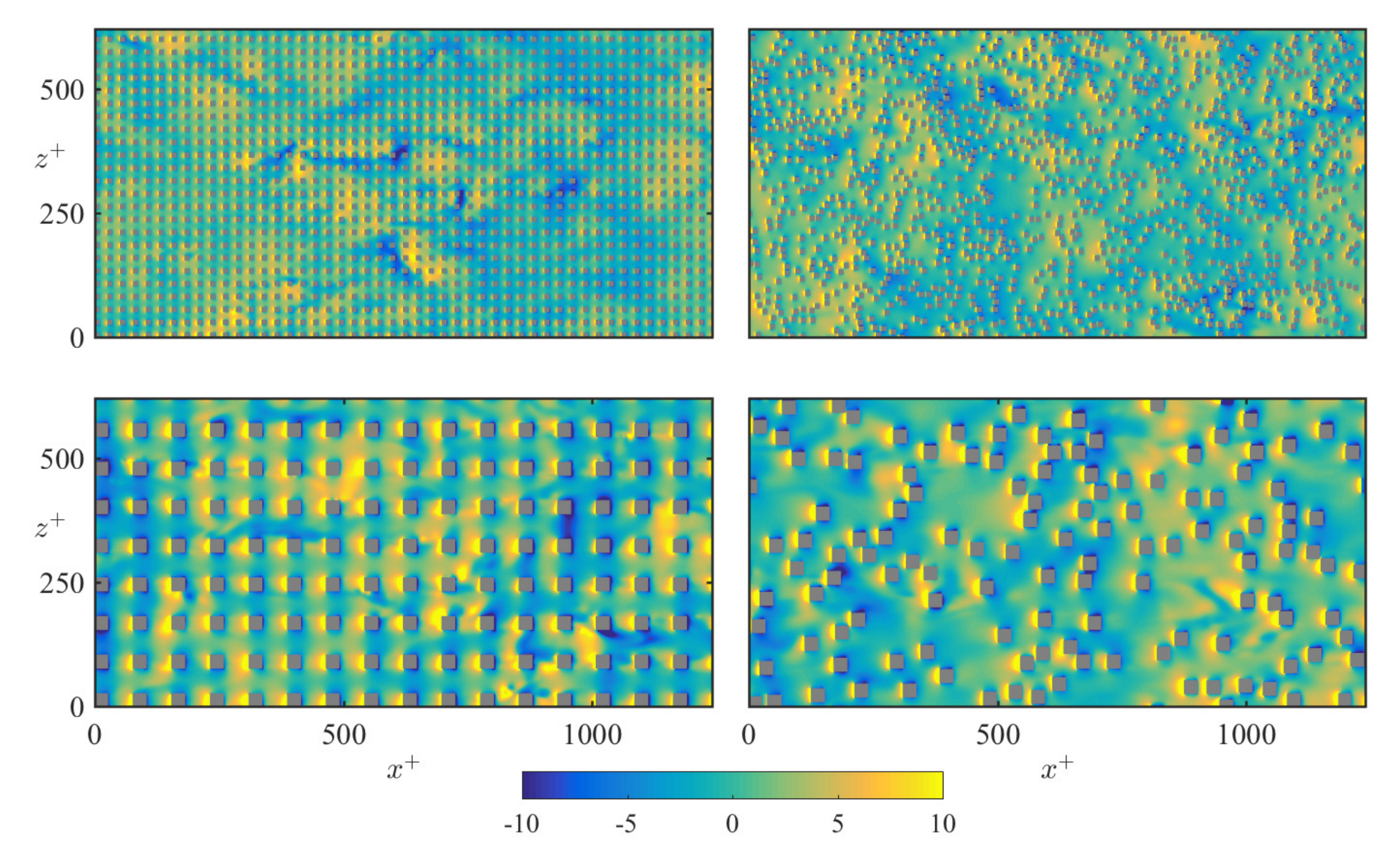}
\mylab{-168mm}{104mm}{(\aaa)}
\mylab{-88mm}{104mm}{(\bbb)}
\mylab{-171mm}{58mm}{(\ccc)}
\mylab{-91mm}{58mm}{(\ddd)}
}
\caption{  
Instantaneous wall pressure snapshots, $p'^+_0=p_0'/(\rho u_\tau^2)$, on superhydrophobic surfaces with texture size $L^+_n=w^+/\sqrt{\phi_s}\approx 26$ with (\aaa) aligned posts and (\bbb) randomly distributed posts, R26$_{1}$, $L^+_n\approx 77$ with (\ccc) aligned posts and (\ddd) randomly distributed posts, R77$_{1}$. From blue to yellow, color shows pressure fluctuations from -10 to 10. 
}
\la{fig:p_rms}
\end{figure}

Next, we investigate the effect of geometric randomness on the interfacial robustness of the gas-liquid interface by investigating wall pressure fluctuations and reconstructed interface deformations. Figure \ref{fig:p_rms} shows instantaneous wall pressure fluctuations for both periodically aligned posts and randomly distributed posts. For both cases, we found that pressure signals show two distinct features: a time-fluctuating signal due to the time variation of the overlying turbulence and a time-independent signal associated with stagnation of the slipping flow near the leading edge of the posts. Therefore, the total turbulence pressure fluctuation signal, $p'$, is decomposed into a temporally stationary signal, $\tilde{p}$, and a random signal, $p''$, so that

\begin{equation}
p'(x,z,y,t) =\tilde{p}(x,z,y) +p''(x,z,y,t), 
\label{eq:e92}
\end{equation} 
where $\tilde{p}$ is averaged over time. The root-mean-square (rms) wall pressure fluctuations of the randomly distributed post, $\tilde{p}^+_{rms_{0}}=\sqrt{ {\frac{\sum_x \sum_z{(\tilde{p}^+_{y=0}})^2}{N_xN_z}}}$, is found to be close to that of the aligned post cases as shown in figure \ref{fig:selfsim}(\aaa). When we consider $\tilde{p}^+_{rms_{0}}$ against slip velocity, $U^+_s$, as shown in figure \ref{fig:selfsim}(\bbb), however, the wall stagnation pressure fluctuation of the randomly distributed posts is larger than that of the aligned posts at a given slip velocity due to about $30\%$ reduction of slip length at the same nominal texture size. The wall stagnation pressure fluctuation increases linearly with increasing slip velocity as 
\beq
\tilde{p}_{rms_{0}}^+ \approx 0.48U_s^+ + 0.79. 
\eeq
The time averaged pressure field is spatially averaged over each unit post. We use the Voronoi diagram to define the effective area surrounding each post. As an example, a Voronoi diagram for RP38 case is shown in figure \ref{fig:config}. The averaged pressure in this way is denoted as $\hat{p}(x,z,y)$. The wall stagnation pressure distribution is found to be self-similar when distance is scaled with $w^+$ and the stagnation pressure is scaled with $\tilde{p}_{rms_{0}}$ as shown in figure \ref{fig:selfsim}(\ccc). 

\begin{figure}
\centerline{
\myfig{166mm}{!}{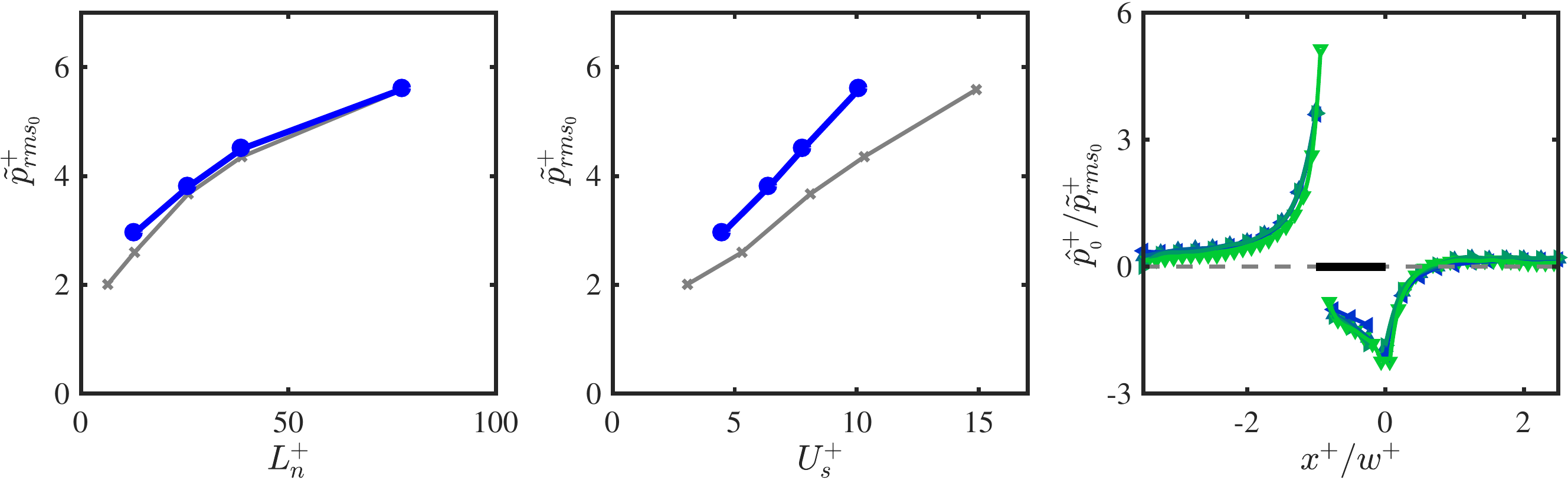}
\mylab{-167mm}{48mm}{(\aaa)}
\mylab{-112mm}{48mm}{(\bbb)}
\mylab{-59mm}{48mm}{(\ccc)}
}
\caption[]{  (\aaa) Wall rms pressure fluctuation due to stagnation, $\tilde{p}^+_{rms_{0}}$ against nominal texture size $L^+_n$; (\bbb) $\tilde{p}^+_{rms_{0}}$ versus slip velocity $U_s^+$.  \linesolidcirc: randomly distributed posts; \linecross: aligned posts. (\ccc) Self-similar profiles of wall pressure distribution due to stagnation on the gas-liquid interface. The plots show pressure profiles on the centerline going through the middle of post width. $L^+_n\approx13$; \linesolidtri: $L^+_n\approx26$;  \linesolidrtri: $L^+_n\approx38$; \linesoliddtri: $L^+_n\approx77$. 
}
\la{fig:selfsim}
\end{figure}

\begin{figure}
\centerline{
\myfig{140mm}{!}{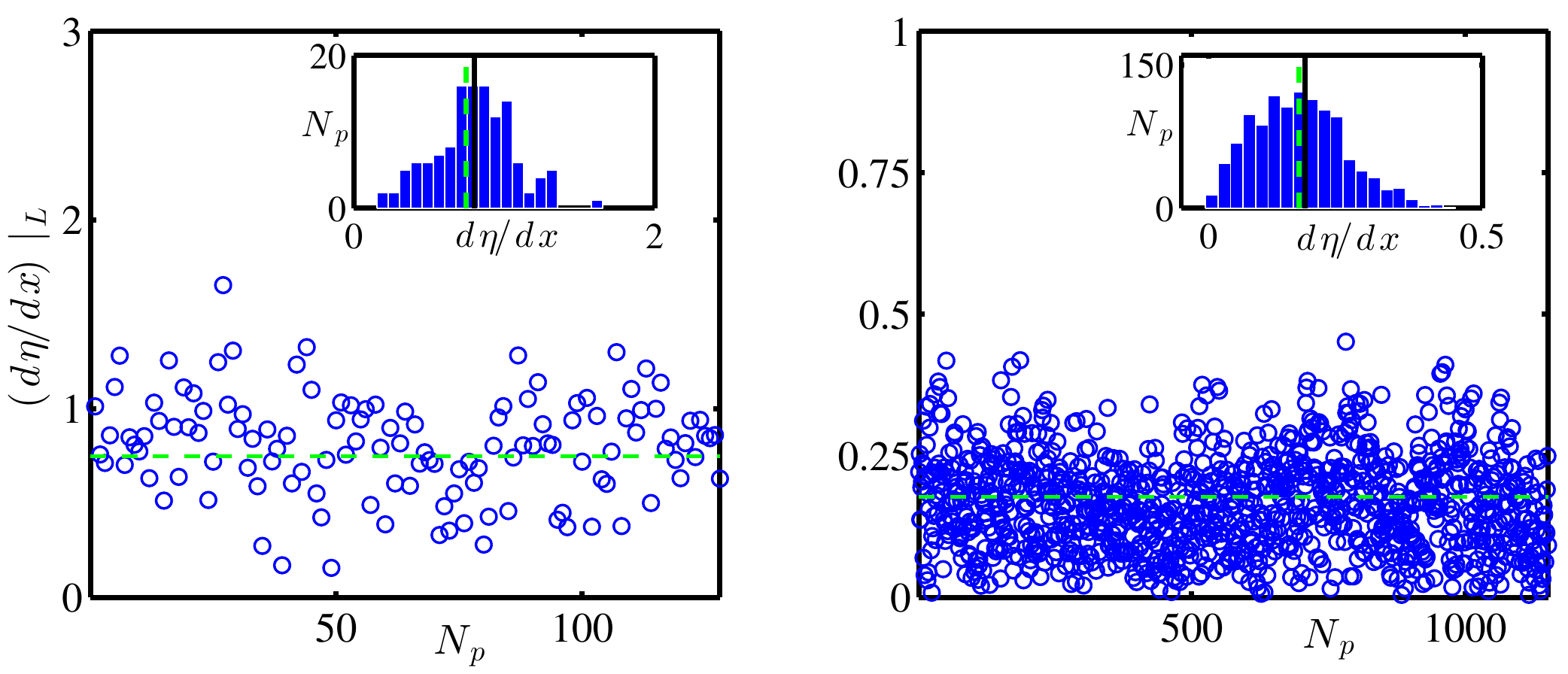}
\mylab{-143mm}{58mm}{(\aaa)}
\mylab{-70mm}{58mm}{(\bbb)}
}
\caption[]{  
Interface deformation angle, $(d\eta/dx|_L)$, due to stagnation pressure for randomly distributed posts versus number of posts, $N_p$, as compared to aligned posts with Weber number $We^+={5\times 10^{-3}}$ for texture size (\aaa) $L_n^+\approx77$; (\bbb) $L_n^+\approx26$. \circle: Randomly distributed posts; Insets: histogram of interface deformation angle due to stagnation pressure for randomly distributed posts versus number of posts. \solid: mean of $(d\eta/dx|_L)$ of randomly distributed posts; \dashed: $(d\eta/dx|_L)$ of aligned posts.
}
\la{fig:maxslope}
\end{figure}

More detailed analysis on the deformation of the interface reveals that the randomness of the textures decreases the stability of gas-liquid interface. 
From time-averaged wall pressure fields, we obtain interface deformation by solving the Young-Laplace equation, Eq. (\ref{eq:Y-L}). We analyze the interface stability in terms of maximum interface deformation for each individual post in different locations. The maximum deformation angle occurs at the leading edge of posts facing upstream slipping flow. The contact angle at the leading edge is defined as $\theta_L=\frac{\pi}{2} + \text{tan}^{-1}({d\eta}/{dx}|_{L})$, where $(d\eta/dx)_{L}$ is the slope of the interface profile at the leading edge \citep{Seo2015}. The onset of interface breakage can be triggered when the large deformation angle of interface exceeds the advancing contact angle that can be achievable only with chemical coating, $\theta_L\gtrapprox 120 ^\circ$. In figure \ref{fig:maxslope}, the deformation angles are reported for each randomly distributed post. Due to randomness in the pattern, the deformation angle for each post is different depending on the arrangement of the texture. The maximum deformation angle is more than twice that of aligned posts at matched $L^+_n\approx77$ as shown in figure \ref{fig:maxslope}(\aaa). For smaller texture size $L^+_n\approx26$, the maximum deformation angle is 2.6 times that of aligned posts. The distributions of deformation angles plotted in the inset of figure \ref{fig:maxslope} show that the peak of the distribution is near the mean, and the mean of deformation angle for randomly distributed posts is close to that of aligned posts. Throughout all texture sizes, $L^+_n\approx13-77$, the maximum deformation angles of randomly distributed posts are plotted in figure \ref{fig:maxslope_L}(\aaa). The maximum deformations of SHS with randomly distributed textures are more than twice that of the periodically aligned post cases. 

Based on our findings, we provide the critical texture size that limits the stable operation of SHSs in turbulent flows in figure \ref{fig:maxslope_L}(\bbb). Due to the increases maximum deformation, the critical texture size of SHS with randomly distributed textures is less than that of aligned posts. 
The maximum allowed $L^+$ is reduced by factor of $\approx 0.5$, compared to aligned case for $We^+=10^{-3}-10^{-2}$.  

\begin{figure}
\centerline{
\myfig{140mm}{!}{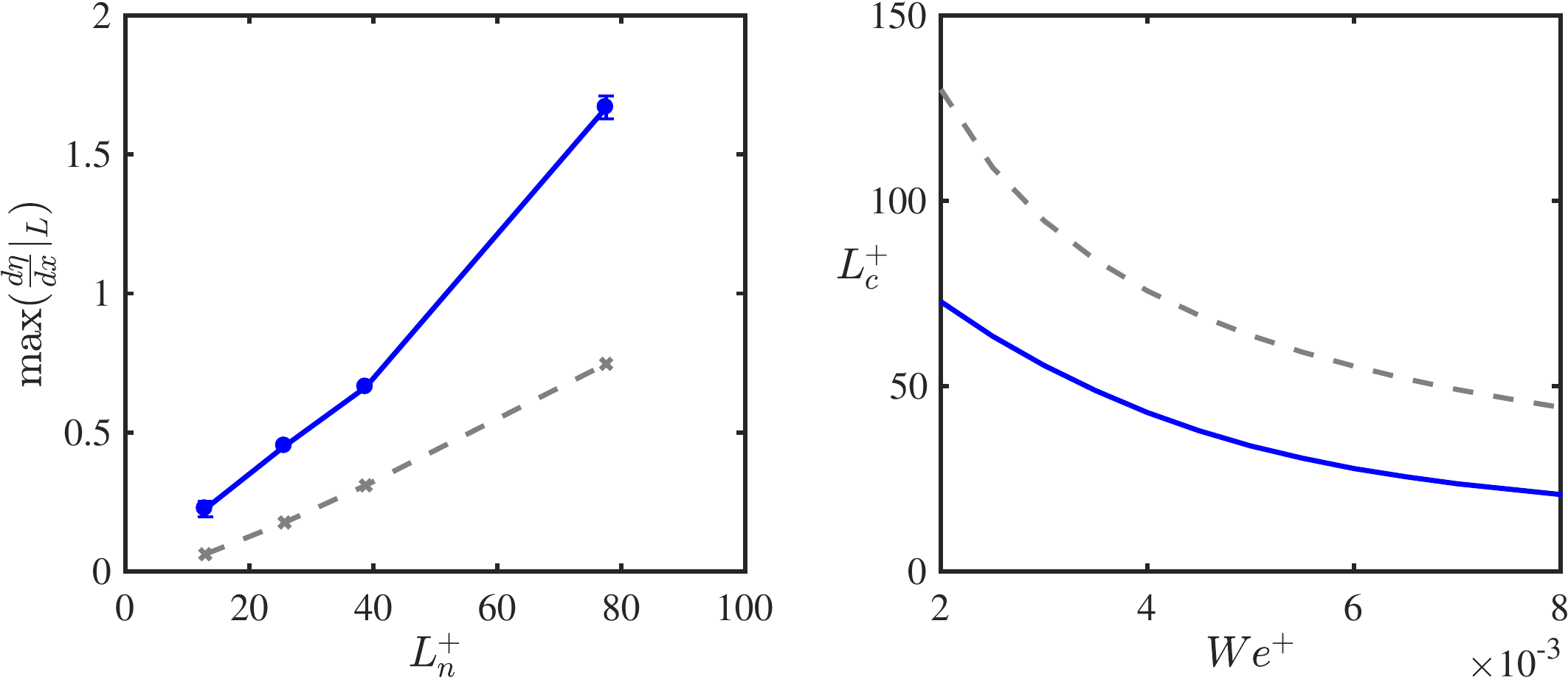}
\mylab{-142mm}{60mm}{(\aaa)}
\mylab{-71mm}{60mm}{(\bbb)}
}
\caption[]{  
(\aaa) Comparison of maximum interface deformation slope, max $(d\eta/dx)|_{L}$, due to stagnation pressure for randomly distributed posts and aligned posts versus effective texture size, $L^+_n=w^+/\sqrt{\phi_s}$, at Weber number $We^+= {5\times 10^{-3}}$. \linesolidcirc: randomly distributed posts; \linecross: aligned posts \citep{Seo2015}. For $L^+_n\approx13$ and $L^+_n\approx 77$, error bars are plotted by taking account for different texture arrangements. (\bbb) The critical texture size, $L^+_c$, is plotted against $We^+$ when $\theta_{adv}=120^{\circ}$. \solid: current study, \dashed: aligned posts.  
}
\la{fig:maxslope_L}
\end{figure}

\section{Summary}
\la{sec:conclusions}

The present work studied the effects of randomness of the SHS texture distribution on hydrodynamic performance when exposed to an overlaying turbulent flow. We conduct direct numerical simulations of turbulent flows over randomly patterned interfaces, considering texture size $w^+\approx 4-26$, and solid fractions, $\phi_s=11-25\%$. Patterned slip and no-slip boundary conditions are imposed on channel walls modeling the entrapped gas within the SHS roughness. The slip length of random textures matches with a linear prediction from Stokes solution in the limit of small texture size, $w^+\approx 4$. For larger textures this linear growth slows down in a fashion similar to previous observations in aligned textures. In the small texture size limit, the solid fraction dependency of the slip length from the DNS data indicates a good agreement with the trends suggested by the Stokes flow solutions. The slip length of random textures is found to be about $30\%$ less than that of aligned textures keeping $Re_\tau$, $w^+$, and $\phi_s$ the same. We show that for fixed texture size, solid fraction, and wall friction, the randomness of the feature distribution decreases the stability of the gas pocket as compared to surfaces with aligned features. The rms wall pressure fluctuation of randomly distributed posts is close to that of aligned posts for fixed wall shear. 
When plotted against the slip length, the pressure fluctuations of the random surface is larger than that for a surface with aligned textures. 
The deformation of the gas-liquid interface is obtained through Young-Laplace equation and the interfacial stability is assessed using the maximum deformation angle at the leading edge of the posts. The detailed analysis on the maximum deformation of individual post sites shows that the randomness of texture arrangement increases the possibility of interface breakage as compared to the aligned case, this is because, at worst, the post encounters interface deformation angles at least twice as large as those in the case of aligned posts. 

This work was supported by the Office of Naval Research under grant 3002451214. 
The program manage is Dr. Ki-Han Kim. 
The authors greatly appreciate the Kwanjeong Educational Foundation for the funding support for Jongmin Seo. 

\bibliographystyle{unsrt}
\bibliography{PRF17Random.bib}

\end{document}